\documentclass[12pt]{article}
\usepackage{times}
\usepackage{amsmath,amssymb,graphicx}
\usepackage[margin=1in]{geometry}
\usepackage{textcomp}

\begin{document}

\begin{center}
{\LARGE \bf Quantum Saturation of Magnetoelectric Coupling in Fe$_3$O$_4$ Nanoparticles} \\[1em]
Jian Huang$^{1}$, Fatemeh Aghabozorgi$^{2}$, and Stephanie Brock$^{2}$ \\[0.5em]
$^{1}$Department of Physics and Astronomy, Wayne State University, Detroit, MI, USA \\[0.2em]
$^{2}$Department of Chemistry, Wayne State University, Detroit, MI, USA \\[1em]
\end{center}

\begin{abstract}

We report magnetoelectric coupling in nanoparticle assemblies that persists to temperatures over 300 times lower than in previous studies. The ME response saturates at low temperature, revealing a quantum plateau of the coupling. A field dependence analysis shows a crossover from quadratic to linear behavior, captured by a phenomenological expansion $C(B,T) \simeq C_0(T) + a_1(T) B + a_2(T) B^2$. The magnitude of the extracted quadratic coefficient, $|a_2(T)|$, follows a power law $|a_2(T)| \sim T^{-\alpha}$ with $\alpha \approx 1.15$, indicating proximity to a quantum critical regime. The observed saturation reflects a new intrinsic energy scale, distinct from finite-size or extrinsic effects. These results establish nanoparticle assemblies as a new platform for studying quantum magnetoelectric phenomena.
\end{abstract}

\section*{Introduction}

The coupling between magnetic and electric orders in quantum materials has emerged as a defining theme in condensed matter physics, bridging fundamental questions of many-body entanglement with opportunities for next-generation quantum technologies~\cite{spaldin2005renaissance,ramesh2007multiferroics,fiebig2016evolution}. Magnetoelectric (ME) coupling, where magnetic moments and electric polarization are intertwined, was first predicted and observed in Cr$_2$O$_3$~\cite{Dzyaloshinskii1960,Astrov1961} and is now known to typically involve classical interactions between spin, lattice, and charge degrees of freedom~\cite{Fiebig2005,kimura2003magnetic,cheong2007multiferroics,tokura2014multiferroics}. However, recent theoretical work has predicted that quantum fluctuations can fundamentally alter this picture, leading to multiferroic quantum criticality where both magnetic and ferroelectric quantum critical points coexist~\cite{Narayan2019}.

The experimental exploration of quantum ME phenomena has been severely limited by the energy scales involved. Most studies of ME coupling operate at temperatures above 4.2 K, with corresponding energies $k_B T > 300~\mu$eV, well above the regime where quantum many-body effects dominate~\cite{rado1975electric,Rado1977_LinearBilinear}. Studies at the true quantum limit~\cite{Rowley2014,Narayan2019,Eerenstein2006,fiebig2016evolution}, where thermal fluctuations become negligible compared to quantum fluctuations, have remained experimentally inaccessible.

Iron oxide nanoparticles present a unique opportunity to access quantum ME states. Unlike bulk magnetite, which exhibits robust ferrimagnetic order and a well-defined Verwey transition~\cite{Walz2002,Garcia2004}, nanoscale confinement enhances the surface-to-volume ratio~\cite{Yoo2016,Wang2020_ColossalMENP,Nguyen2021_Fe3O4_NP_Review,Song2022_CoreShellMENP}, modifying spin-orbit coupling at interfaces~\cite{Batlle2002,Kodama1996,Martinez2022} and enabling quantum tunneling of magnetization. Furthermore, finite-size effects disrupt long-range magnetic ordering, while geometric frustration from surface spin disorder and competing exchange interactions creates conditions favorable for emergent quantum states. Such frustrated spin systems can host collective quantum phenomena~\cite{anderson1973resonating,Balents2010} that may support unconventional ME responses.

We report the observation of quantum many-body magnetoelectric states in Fe$_3$O$_4$ nanoparticles using ultralow-temperature capacitance measurements. A micro-cavity platform with 0.01~fF resolution enables quantum-regime measurements down to 6~mK with only 0.1~mg of material. Two striking phenomena emerge: (1)~a sharp ME coupling enhancement at 160~mK ($\approx 14~\mu$eV), signaling quantum-correlated excitations, and (2)~saturation below 40~mK, suggesting a many-body ground state. These energy scales and temperature dependences indicate quantum multiferroic behavior from frustrated magnetic interactions in nanoscale systems.

\section*{Methods} \noindent\textbf{Samples and Ultra-sensitive Capacitance Measurements:} The sample consisted of approximately 0.1 mg of Fe$_3$O$_4$ nanoparticles (NPs) with an average diameter of $20.9 \pm 3.9$~nm. The NPs were sandwiched between two n-doped silicon substrates coated with 90 nm of thermally grown silicon dioxide, forming a parallel-plate capacitor geometry with a gap of $\sim$40-50 $\mu$m between the n$^+$ electrodes (Fig. 1a). This assembly was mounted inside a gold-coated oxygen-free copper microcavity, which provided electromagnetic shielding and thermal anchoring to the mixing chamber plate via a heat sink (Fig. 1b). Electrical connections to the n$^+$ electrodes utilized coaxial feedthroughs with careful impedance matching to minimize parasitic capacitance and noise (Fig. 1c). The ME response was probed by monitoring variations in the dielectric constant through high-precision capacitance measurements with a resolution of $\sim 0.01$ fF. All experiments were conducted in a dilution refrigerator with a base temperature of $\sim$6 mK and applied magnetic fields up to $\pm 0.6$ T. Temperature-dependent measurements employed controlled thermal cycles with sufficient equilibration times. Magnetic field sweeps were performed at fixed temperatures with sweep rates $<0.1$ mT/s. 

\begin{figure}[htbp] 
\centering 
\includegraphics[width=0.8\textwidth]{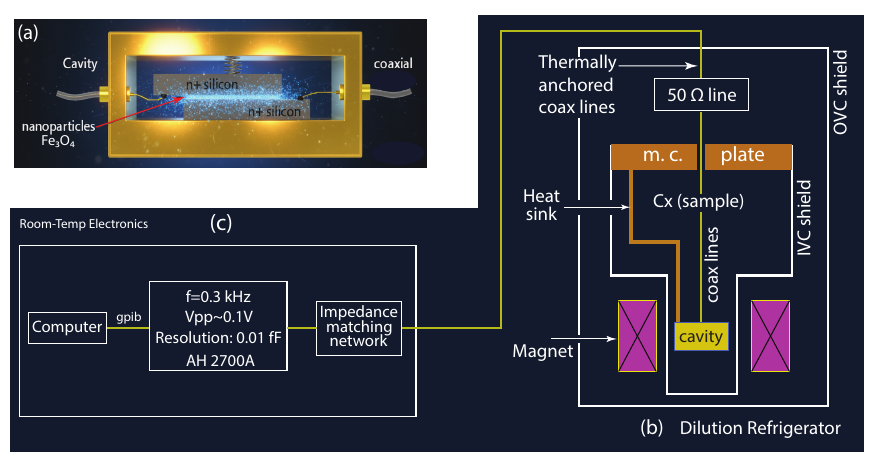} \caption{Figure 1: Ultra-sensitive ME measurement setup and Fe$_3$O$_4$ nanoparticle characterization. \textbf{a.} Micro-cavity assembly containing 0.1 mg Fe$_3$O$_4$ nanoparticles between two n$^+$ Si/SiO$_2$ substrates with a 40-50 $\mu$m gap. Magnetic field is applied perpendicular to the capacitor plates. \textbf{b.} Coaxial wiring and cooling schematics inside the dilution refrigerator (DL); the cavity sits in the bore of a magnet, heat-sinked to the mixing chamber plate. \textbf{c.} Measurement circuit showing the impedance-matched signal path from the low-temperature sample to room-temperature electronics, using an AH2700A bridge that achieves 0.01 fF resolution at 0.3 kHz.} \label{fig:Fig1} 
\end{figure} 

The measured capacitance is directly related to the effective dielectric constant of the nanoparticle layer, $C(B,T) = \varepsilon_{\mathrm{eff}}(B,T) A/d$, where $A$ and $d$ are the electrode area and separation, respectively (see Supplemental for derivation). 
$\varepsilon_{\mathrm{eff}}$ is the effective permittivity. The ME coupling modifies $\varepsilon_{\mathrm{eff}}$ through the expansion of polarization in powers of the magnetic field: $P_i(B,T) = \alpha_{ij}(T) B_j + \beta_{ijk}(T) B_j B_k + \cdots$, where $\alpha_{ij}$ is the linear magnetoelectric tensor and $\beta_{ijk}$ is the quadratic tensor. These terms renormalize the dielectric susceptibility as \[ \varepsilon_{\mathrm{eff}}(B,T) = \varepsilon_0 + \chi_E(T) + \Delta\varepsilon(B,T), \] where $\varepsilon_0$ is the vacuum permittivity, $\chi_E(T)$ is the zero-field dielectric susceptibility of the nanoparticle–substrate assembly, and $\Delta\varepsilon(B,T)$ encodes the field-dependent corrections. The latter contains odd-in-$B$ contributions from the linear magnetoelectric tensor $\alpha_{ij}$ and even-in-$B$ contributions from the quadratic tensor $\beta_{ijk}$. This leads to the leading-order polynomial expansion for the measurable capacitance: 
\begin{equation} 
C(B,T) \simeq C_0(T) + a_1(T) B + a_2(T) B^2. \end{equation} 

To isolate the symmetry-distinct ME channels we first decomposed each field sweep into even and odd components, $C_{\rm even}(B)=\tfrac{1}{2}[C(B)+C(-B)]$ and $C_{\rm odd}(B)=\tfrac{1}{2}[C(B)-C(-B)]$. The leading even response was obtained by fitting $C_{\rm even}(B)=C_0+a_2 B^2$ within a symmetric low-field window $|B|\le B_{\rm max}$. $B_{\rm max}$ was chosen objectively as the largest field for which (i) the residuals of the quadratic fit display no systematic curvature and (ii) the fitted coefficient $a_2$ is stable within one standard error upon small variations of $B_{\rm max}$. Note that, in principle, $a_1$ averages to zero for powders, but we retain it in the fit to capture small residual asymmetries. Higher-order even terms were tested and found not to be statistically significant within the selected window (as the reduction in $\chi^2$ was insignificant). To confirm that the observed quadratic ME response is intrinsic to Fe$_3$O$_4$ and not an artifact of the measurement apparatus, control measurements on Gd$_2$O$_3$ nanoparticles (which lack a ME phase) were performed. These controls show no significant field dependence (see Supplemental Fig.~3), verifying the intrinsic nature of the effect. 

\vspace{10 pt}
\noindent\textbf{Sample Preparation and Characterization:} Fe$_3$O$_4$ nanoparticles were synthesized using thermal decomposition methods (see Supplemental Information for details). Magnetic characterization via SQUID magnetometry revealed superparamagnetic behavior, with zero-field-cooled (ZFC) and field-cooled (FC) magnetization curves diverging around 350 K and a ZFC maximum at $T_b \approx 60$ K. This indicates a broad distribution of blocking temperatures, consistent with the nanoparticle size distribution (see Supplemental Fig. 4). Low-temperature hysteresis loops (not shown) exhibited a strong temperature dependence of coercivity ($H_c = 168$ Oe at 10 K vs. $60$ Oe at 110 K), confirming that magnetic anisotropy energies are comparable to thermal energies in our measurement range. This energy scale competition enables quantum fluctuations to rival classical magnetic switching mechanisms. The reduced saturation magnetization and broad blocking temperature distribution provide the surface spin disorder and magnetic frustration necessary for stabilizing quantum many-body states, which are absent in bulk magnetite. 

\section*{Results} 

\noindent\textbf{Magnetoelectric field dependence} The magnetic field dependence of capacitance (Fig.~2a) provides a direct probe of ME susceptibility and reveals distinct coupling mechanisms across different temperature regimes. At $T=77$ mK, the response is symmetric to leading order and well described by a quadratic dependence, $\Delta C \propto B^2$, characteristic of enhanced second-order ME coupling. To isolate the intrinsic even-in-field contribution, we fit the data with $C_{\rm even}(B)=C_0+a_1B+a_2B^2$ over the restricted low-field range $|B|\leq 0.2$~T (see Methods for objective selection). The fit (shown in Fig. 2b) captures the data with excellent agreement ($R^2 = 0.998$), yielding $a_2 = -0.593 \pm 0.013$, $a_1 = (0.0 \pm 2.3)\times 10^{-4}$ (consistent with zero), and $C_0 = 53.7657 \pm 0.0003$. At higher fields, the capacitance deviates systematically from a simple quadratic form, reflecting nonlinear magnetization and magnetostrictive contributions; these high-field deviations were excluded from the low-field analysis. The dominance of the quadratic term highlights the role of quantum spin–lattice–charge correlations in governing the dielectric susceptibility in this regime. 

\begin{figure}[htbp] 
\center 
\vspace{0pt} 
\includegraphics[width=0.95\textwidth]{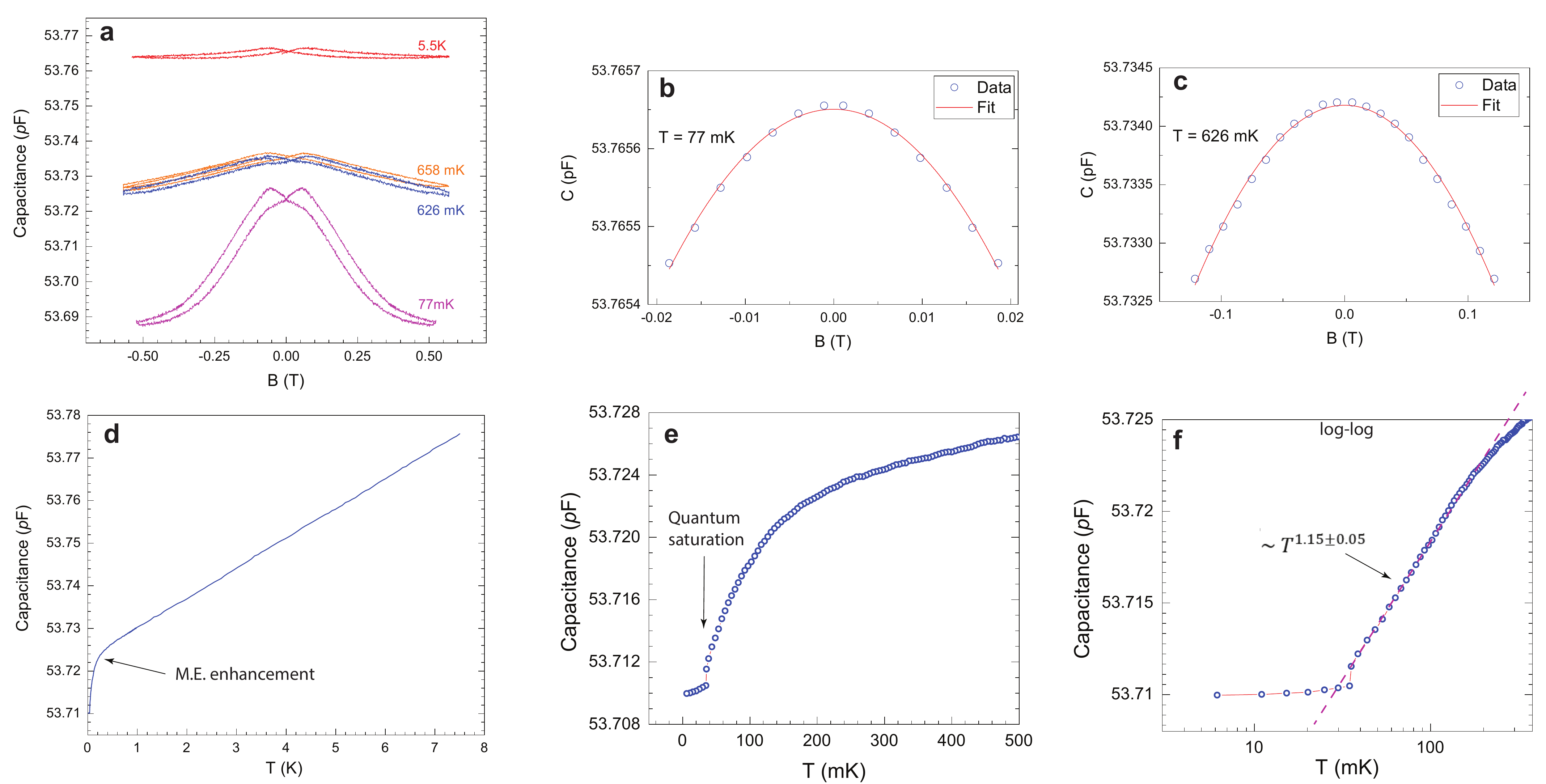} \vspace{0pt} 
\caption{\label{fig:Fig2}\textbf{Quantum ME coupling and critical scaling.} \textbf{a.} B-field dependence of capacitance at various temperatures showing the crossover from classical (5.5~K, field-independent) to quantum (77~mK) ME response. \textbf{b.} Polynomial fit for $C_{even}(B)$ at 77~mK. \textbf{c.} Polynomial fit for $C_{even}(B)$ at 626~mK. \textbf{d.} Temperature dependence of capacitance in zero field, highlighting a transition from weakly varying classical behavior above 1 K to a pronounced enhancement below $T^\ast \approx 160$ mK. \textbf{e.} Detailed view of quantum enhancement region revealing sharp increase below $T^*$ followed by abrupt saturation below $T_{\rm sat}$, indicating formation of a quantum many-body ground state. \textbf{f.} Log-log analysis of the enhancement region showing power-law scaling $\Delta C \propto T^{1.15 \pm 0.05}$ above $T_{sat}$. } 
\end{figure}

We note that the fitted linear coefficient $a_1$ is negligible at low temperatures, consistent with the random orientations of the nanoparticles in the cavity. A linear ME effect requires a fixed tensorial coupling $\alpha_{ij}$, which averages to zero over an ensemble of randomly oriented particles. By contrast, the quadratic response arises from higher-order couplings $\beta_{ijkl}$ that survive orientational averaging, yielding the observed $\Delta C \propto B^2$ dependence. At elevated temperatures (e.g., 650 mK), a weak residual odd-in-field contribution is detected, which we attribute to incomplete orientational cancellation or surface-spin imbalance effects. 

At intermediate temperature of $T=626$~mK, the field dependence of the capacitance remains predominantly quadratic, $C(B)\approx C_0 + a_2 B^2$, but with a substantially reduced curvature compared to the 77~mK data (Fig.2c). The extracted quadratic coefficient $a_2=-0.1046 \pm 0.0017$ is nearly six times smaller in magnitude than at 77~mK, indicating that the enhancement of ME susceptibility is rapidly suppressed upon heating. A small odd-in-field term, $a_1=-1.78\times 10^{-5} \pm 1\times 10^{-4}$, is also present, consistent with a weak breaking of inversion or time-reversal symmetry, possibly due to surface spin imbalance in the nanoparticles. The overall fit quality ($R^2 \approx 0.985$) suggests that additional higher-order contributions become relevant at this temperature, reflecting increased thermal fluctuations that reduce the dominance of the purely quadratic response. 

At $T=5.5\ \mathrm{K}$, the field dependence is not well described by a simple quadratic form. Even/odd decomposition shows that the even component, $C_{\rm even}(B)$, departs systematically from $B^2$ at moderate fields, and attempts to fit higher-order polynomials lead to significant residuals. This behavior is consistent with the dominance of thermally activated magnetization~\cite{Bean1959,Dormann1997} and magnetostrictive processes~\cite{Fiebig2005,rado1975electric,Eerenstein2006,Yoo2016,Buschow1977,Battle2002} at elevated temperatures, marking a crossover from the low-temperature regime where a quadratic, second-order ME response prevails.

The temperature dependence of the capacitance in zero magnetic field (Fig.~2d) shows a qualitative change in ME response as the system enters the quantum regime. Above 1~K the capacitance is nearly temperature independent, but below $T^* \approx 160$~mK ($k_B T^* \approx 14$~$\mu$eV) a sharp enhancement appears, with the capacitance increasing by more than 60~fF before saturating abruptly at $T_{\mathrm{sat}} \approx 40$~mK (Fig.~2e). The onset scale of 14~$\mu$eV is one to two orders of magnitude smaller than typical single-particle magnetic excitations in iron oxides ($\sim 1$~meV), suggesting that the anomaly originates from collective dynamics rather than isolated spin flips or phonon modes. 

To quantify this regime, we subtracted the low-temperature baseline capacitance and analyzed the excess response, $\Delta C(T) = C(T) - C_{\mathrm{sat}}$. As shown in Fig.~2f, $\Delta C$ follows a near-linear power law, \[ \Delta C(T) \propto T^{\alpha}, \qquad \alpha = 1.15 \pm 0.05, \] over the range $40~\text{mK} \lesssim T \lesssim 200~\text{mK}$. This form is inconsistent with an activated behavior, $\Delta C(T) \propto e^{-\Delta/T}$, which would yield exponential suppression, and also deviates from a Curie–Weiss law expected for classical fluctuations. Instead, the power-law scaling with $\alpha\sim1$ points to a quantum-critical regime governed by soft collective modes~\cite{Sachdev2011}. The agreement between the full capacitance curve (Fig.~2a) and the subtracted response (Fig.~2d,f) thus establishes a crossover from a classical regime at high temperatures to a quantum-critical regime below $T^*$, with the saturation at $T_{\mathrm{sat}}$ marking a distinct low-temperature state. 

The abrupt saturation observed in Fig.~2e represents a distinct feature at ultralow-temperature: below $T_{\mathrm{sat}} \approx 40$~mK ($k_B T_{\mathrm{sat}} \approx 3.5~\mu$eV), the capacitance levels off. It remains nearly constant down to the base temperature of $\sim 6$~mK. This plateau indicates that the divergent growth of the susceptibility seen at higher temperatures is cut off, consistent with the stabilization of a distinct low-energy state. 
One natural interpretation is that quantum fluctuations become exhausted as the system approaches its ground state, such that further cooling no longer enhances spin–lattice–charge correlations. Alternatively, finite-size effects or surface states intrinsic to the nanoparticles may impose a cutoff that mimics saturation. In either case, the emergence of a sharp low-temperature plateau is unexpected within conventional models of ME coupling and points to a regime where new collective physics must be considered~\cite{Narayan2019,Rowley2014,Millis1993}. 

\section*{Discussion} 

The capacitance measurements establish a set of stringent empirical constraints. In zero magnetic field, the response shows a sharp enhancement below $T^\ast \simeq 160$~mK ($k_B T^\ast \approx 14~\mu$eV) and an abrupt saturation below $T_{\rm sat} \simeq 40$~mK ($k_B T_{\rm sat} \approx 3.5~\mu$eV) (Fig.~2d,e). Analysis of the excess response, $\Delta C(T) = C(T) - C_{\rm sat}$, yields a power-law dependence $\Delta C \propto T^{1.15 \pm 0.05}$ over the range $40$–$200$~mK (Fig.~2d), consistent with gapless or nearly-gapless collective fluctuations. Field-dependent measurements provide complementary evidence: the ME response, is dominated by even-in-B within error, with small residual odd terms at intermediate T. However, the strength of this quadratic coefficient $a_2(T)$ undergoes a dramatic change—it is strong at the lowest temperatures (77 mK) but becomes vanishingly small at higher temperatures (5.5 K) (Fig.~2a). Together, these results indicate a continuous evolution from a regime of strong quantum-mechanical, second-order ME coupling at ultralow $T$ to a state where these correlations are thermally suppressed. 

A natural interpretation is that the enhancement below $T^\ast$ arises from the thermal population of a soft collective mode in which spin, lattice, and charge degrees of freedom are strongly coupled. The near-linear power-law scaling~\cite{Rowley2014} suggests that the dielectric susceptibility is governed by quantum-critical fluctuations of this mode, while the saturation at $T_{\rm sat}$ marks either the opening of a small gap or a cutoff imposed by finite-size or surface effects. The energy scales involved, in the few-$\mu$eV range, are far smaller than single-particle excitations in iron oxides, pointing to a many-body origin such as hybrid spin–phonon modes or exchange-striction–driven fluctuations strongly renormalized by nanoscale disorder. This distinguishes the observed quantum-critical regime from the robust ME coupling achieved at room temperature in classically ordered nanoparticle heterostructures~\cite{Wang2019_IronOxideBaTiO3}, highlighting the fundamentally different nature of fluctuations in the quantum limit. 

Alternative explanations must also be considered. A cooperative super-spin-glass freezing~\cite{Jonsson1995,Djurberg1997,Sasaki2005} could mimic a power-law regime and produce an apparent saturation, although such a scenario would be expected to show strong frequency dependence, aging, or hysteresis, which we do not observe within our measurement bandwidth (e.g., 0.3 kHz as mentioned in the Supplemental). Likewise, artifacts from the measurement setup were checked through control tests and found to be insufficient to account for the magnitude of the effect. The phenomenology recalls concepts such as resonating valence bonds in frustrated magnets~\cite{anderson1973resonating,Balents2010,Savary2017}, where no single classical configuration suffices, and the ground state is reminiscent of states described by entangled superpositions. By analogy, the Fe$_3$O$_4$ nanoparticles appear to host collective spin–lattice–charge fluctuations that cannot be factorized into independent modes. While our data do not establish such a state microscopically, the resemblance highlights the possibility that similar many-body principles may govern ME responses in nanoscale oxides. 

Taken together, the evidence favors a collective, soft-mode interpretation in which strongly renormalized spin–lattice–charge fluctuations dominate the ME response in Fe$_3$O$_4$ nanoparticles at ultralow temperatures. The sharp crossover into a saturated state below $T_{\rm sat}$ highlights the existence of a new energy scale, suggesting that the ground state hosts correlations qualitatively different from both classical multiferroic responses and single-particle excitations. Future spectroscopy, calorimetry, and field-tuning studies will be crucial to resolve the microscopic nature of this correlated regime. 

\section*{Conclusions} 

Our capacitance measurements on Fe$_3$O$_4$ nanoparticles reveal an unusual combination of quantum-critical scaling and low-temperature saturation in the ME response. The results cannot be explained by conventional single-particle excitations or classical dielectric models, and instead point to correlated collective fluctuations at ultralow energy scales. While the microscopic nature of these excitations remains to be determined, the robustness of the experimental signatures demonstrates that nanoparticles provide a fertile setting for uncovering novel regimes of coupled spin–lattice–charge dynamics. In the broader context, our findings open a pathway to explore collective quantum states in nanoscale systems, with potential implications ranging from refined models of multiferroic quantum criticality to practical concepts for ultra-low-temperature refrigeration and quantum devices. 

\section*{Acknowledgment} The NSF supported this work under No.~2232489. \newpage 

\bibliographystyle{naturemag} \bibliography{biblio_ME}

\newpage \section*{Supplemental} 

\subsection{Capacitance of the layered structure} \paragraph{Layer model and extraction of nanoparticle permittivity.} The device consists of three dielectric layers in series: SiO$_2$ (top, thickness $d_1$), the nanoparticle film (thickness $d_2$), and SiO$_2$ (bottom, thickness $d_3$). Treating each layer as a parallel-plate capacitor, the total capacitance is \[ C(B,T)=A\left(\frac{d_1}{\varepsilon_0\varepsilon_{r,1}}+\frac{d_2}{\varepsilon_0\varepsilon_{r,2}(B,T)}+\frac{d_3}{\varepsilon_0\varepsilon_{r,3}}\right)^{-1}, \] where $\varepsilon_{r,1}=\varepsilon_{r,3}=3.9$ (SiO$_2$) and $\varepsilon_{r,2}(B,T)$ is the effective relative permittivity of the nanoparticle layer. This relation can be inverted to obtain $\varepsilon_{r,2}(B,T)$ directly from the measured $C(B,T)$: \[ \varepsilon_{r,2}(B,T)=\frac{d_2}{\varepsilon_0\Big(\dfrac{A}{C(B,T)}-\dfrac{d_1}{\varepsilon_0\varepsilon_{r,1}}-\dfrac{d_3}{\varepsilon_0\varepsilon_{r,3}}\Big)}. \] For small changes we linearize to find \[ \Delta C \approx C^{2}\,\frac{d_2}{\varepsilon_0 A}\,\frac{\Delta\varepsilon_{r,2}}{\varepsilon_{r,2}^{2}}, \] which can be inverted to obtain $\Delta\varepsilon_{r,2}$ from the measured $\Delta C$. Because the SiO$_2$ layers are only 90~nm thick while the nanoparticle layer is $40$--$50\ \mu$m thick, their per-area capacitances are $\gtrsim 30\times$ larger than that of the nanoparticle layer for plausible $\varepsilon_{r,2}$ values; thus the total capacitance is dominated by the nanoparticle layer and measured changes in $C$ are dominated by changes in $\varepsilon_{r,2}$. 


\subsection{Calibration of the Capacitance Setup}

The measurement of ME coupling via capacitance requires exceptional sensitivity to changes at the level of attofarads (10$^{-18}$ F). To establish the validity of our measurements and rule out instrumental artifacts, we performed multiple calibration and validation steps. 

First, the measurement bridge was calibrated against standard capacitors at room temperature, and the parasitic capacitance of the coaxial wiring was carefully measured and nulled. Second, the stability and field dependence of the empty capacitor geometry (without nanoparticles) were verified to be negligible compared to the signals reported here. 

Third, to confirm the intrinsic nature of the ME response and rule out frequency-dependent artifacts, control measurements were performed at multiple frequencies (0.3 kHz, 0.5 kHz, and 1 kHz). The absence of significant frequency dependence in both the zero-field capacitance and the field-induced variations demonstrates that the observed effects represent genuine ME coupling rather than electrode polarization or other frequency-dependent artifacts. The 0.3 kHz operating frequency was selected as it provides optimal signal-to-noise ratio while being sufficiently low to avoid complications from finite conductivity in the nanoparticle assembly. The following section details a crucial control experiment performed with a non-magnetoelectric material to further validate our findings.

\begin{figure}[htbp] \vspace{-5pt} \centering \includegraphics[width=0.5\textwidth]{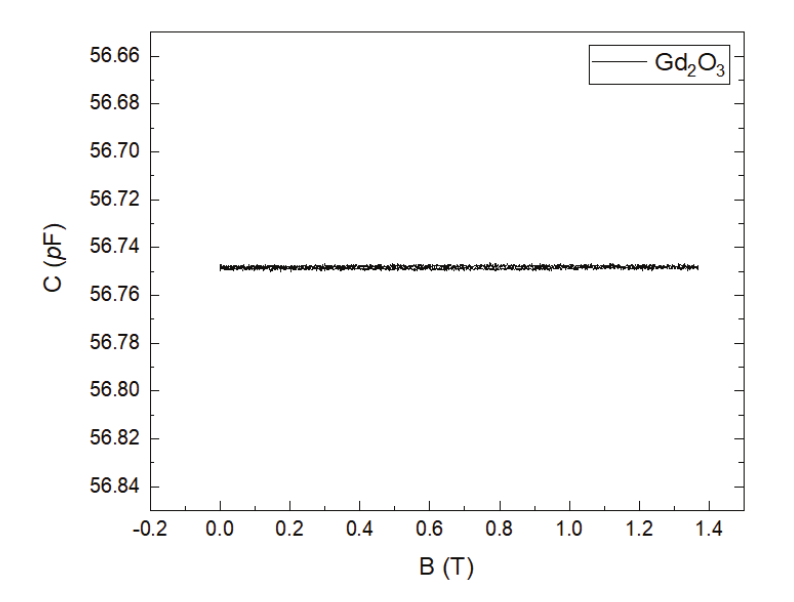} \vspace{-5pt} \caption{\textbf{Control measurement: Magnetic field dependence of capacitance for Gd$_2$O$_3$ nanoparticles.} Data were acquired at 80 mK using the same cavity and electronics as for the Fe$_3$O$_4$ measurements. The capacitance shows no significant field dependence and, critically, no quadratic curvature, confirming that the large $a_2$ coefficient measured in Fe$_3$O$_4$ is an intrinsic effect and not an experimental artifact.} \label{fig:s1_gd2o3} 
\end{figure} 

\subsection{Control Experiment with Gd$_2$O$_3$ Nanoparticles} To confirm that the pronounced quadratic ME response observed in Fe$_3$O$_4$ is an intrinsic material property and not an artifact of the measurement setup (e.g., magnetoresistance of the silicon substrates or the coaxial cables), a control experiment was performed under identical conditions using Gd$_2$O$_3$ nanoparticles. Gd$_2$O$_3$ is a wide-bandgap paramagnetic insulator with no expected ME coupling. As shown in Fig. \ref{fig:s1_gd2o3}, the capacitance of the Gd$_2$O$_3$ sample is flat as a function of magnetic field. A fit to the quadratic form $C(B) = C_0 + a_2 B^2$ yields a negligible coefficient $a_2^{\mathrm{Gd_2O_3}}$ compared to the value of $a_2^{\mathrm{Fe_3O_4}} =-0.59$ fF/T$^2$ measured at 77 mK. This control experiment provides definitive evidence that the reported ME effects are intrinsic to the Fe$_3$O$_4$ nanoparticles. 

\subsection{Synthesis and Characterization of Fe$_3$O$_4$ Nanoparticles} 

Fe$_3$O$_4$ nanoparticles were synthesized via controlled oxidation of an Fe(OH)$_2$ precursor in an aqueous solution. The synthesis follows the reaction sequence: Fe$^{2+}$ + 2OH$^-$ $\rightarrow$ Fe(OH)$_2$, followed by partial oxidation 3Fe(OH)$_2$ + $\frac{1}{2}$O$_2$ $\rightarrow$ Fe(OH)$_2$ + 2FeOOH + H$_2$O, and final condensation Fe(OH)$_2$ + 2FeOOH $\rightarrow$ Fe$_3$O$_4$ + 2H$_2$O. 

\begin{figure}[htbp] 
\centering 
\includegraphics[width=0.85\textwidth]{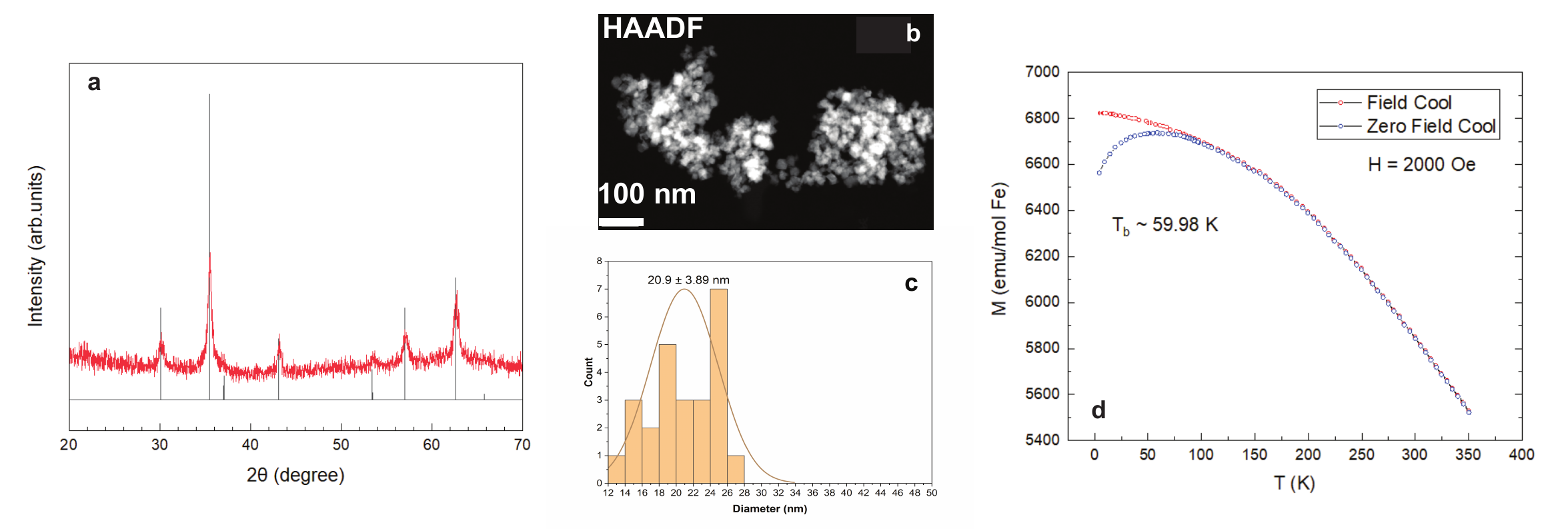} \caption{\textbf{Structural and magnetic characterization of Fe$_3$O$_4$ nanoparticles.} \textbf{a,} X-ray diffraction pattern showing characteristic peaks of the inverse spinel structure at 2$\theta = 30.1^\circ$, $35.5^\circ, 43.1^\circ, 53.4^\circ, 57.0^\circ$, and $62.6^\circ$, corresponding to (220), (311), (400), (422), (511), and (440) reflections, respectively. The well-defined peaks confirm high crystallinity with no detectable impurity phases. Peak broadening analysis yields an average crystallite size of 20--25 nm. \textbf{b,} Transmission electron microscopy image revealing spherical nanoparticles with minimal aggregation. Scale bar: 100 nm. \textbf{c,} Particle size distribution histogram from TEM analysis showing a narrow distribution centered at $20.9 \pm 3.9$ nm (Gaussian fit). \textbf{d,} Zero-field-cooled (ZFC) and field-cooled (FC) magnetization under a 2000 Oe applied field, showing superparamagnetic blocking with a ZFC maximum at $T_b \approx 60$ K. \textbf{e,} Low-temperature hysteresis loops at 10 K and 110 K, demonstrating a strong temperature dependence of coercivity ($H_c = 168$ Oe at 10 K vs. 60 Oe at 110 K). } 
\label{fig:s2_char} 
\end{figure} 

X-ray diffraction confirmed the inverse spinel structure with crystallite sizes of 20--25 nm and no detectable impurity phases, indicating single-crystalline nanoparticles (Fig. \ref{fig:s2_char}a). TEM imaging revealed spherical particles with minimal aggregation (Fig. \ref{fig:s2_char}b), and size distribution analysis showed a narrow profile centered at $20.9 \pm 3.9$ nm (Fig. \ref{fig:s2_char}c), consistent with the XRD results. 

Room-temperature magnetization measurements revealed a saturation magnetization $M_s = 58$ emu/g, significantly reduced from the bulk Fe$_3$O$_4$ value of 92 emu/g. This reduction provides direct evidence for significant surface spin disorder~\cite{sayed2019controlling}, a characteristic of high-surface-area nanoparticles. Magnetic characterization via SQUID magnetometry showed classic superparamagnetic behavior. The ZFC and FC magnetization curves diverge around 350 K, with a ZFC maximum at $T_b \approx 60$ K (Fig. \ref{fig:s2_char}d), indicating a distribution of blocking temperatures. Low-temperature hysteresis loops (Fig. \ref{fig:s2_char}e) demonstrated a strong temperature dependence of coercivity, confirming that magnetic anisotropy energies are comparable to thermal energies in our measurement range. This competition is a prerequisite for enabling quantum fluctuations to rival classical magnetic switching mechanisms. The combination of reduced saturation magnetization, broad blocking temperature distribution, and finite-size effects provides the surface spin disorder and magnetic frustration necessary for stabilizing the quantum many-body states observed in this work. 

\end{document}